\documentclass{kapproc} 

\usepackage{procps} 

\usepackage[dvips]{graphicx}

\upperandlowercase

\kluwerbib 

\begin{document}

\articletitle[Bars from the Inside Out: An HST Study of their Dusty 
Circumnuclear Regions] 
{Bars from the Inside Out: An HST Study of their Dusty Circumnuclear Regions} 

\author{Paul Martini} 
 
\affil{
Harvard-Smithsonian Center for Astrophysics; 
60 Garden Street, MS 20; Cambridge, MA 02138; USA
}

\begin{abstract}
The results of bar-driven mass inflow are directly observable in 
high-resolution {\it HST}\ observations of their circumnuclear regions. 
These observations reveal a wealth of structures dominated by dust lanes, 
often with a spiral-like morphology, and recent star formation. 
Recent work has shown that some of these structures are correlated with the 
presence or absence of a bar. I extend this work with an investigation of 
circumnuclear morphology as a function of bar strength for a sample of 48 
galaxies with both measured bar strengths and ``structure maps'' computed 
from {\it HST}\ images. 
The structure maps for these galaxies, which have projected spatial 
resolutions of 2 -- 15 pc, show that the fraction of galaxies with grand-design 
(GD) circumnuclear dust spirals increases significantly with bar strength, 
while tightly wound dust spirals are only present in the most axisymmetric 
galaxies. 
In the subset of galaxies classified SB(s), SB(rs), or SB(r), GD structure is 
only found at the centers of SB(s) or SB(rs) galaxies, and not SB(r). 
Bar strength measurements of 45 SB(s), SB(rs), and SB(r) galaxies show 
that SB(s) galaxies have the strongest bars, while SB(r) galaxies have the 
weakest bars. 
As SB(s) galaxies are also observed to most commonly possess dust lanes 
along their leading edges, this is further support of a connection between 
GD structure and bar-driven inflow on larger scales. 
There is also a modest increase in the fraction of loosely wound dust spirals 
at later morphological types, and a corresponding decrease in the fraction of 
chaotic structures. This trend may reflect an increase in the fraction of 
galaxies with circumnuclear, gaseous disks. 
The trend appears to reverse at type Scd, where the fraction of 
galaxies with chaotic circumnuclear dust structure increases dramatically, 
although these data are of poorer quality. 
\end{abstract}


\section{Introduction}

Bars are the most effective means of driving gas toward the centers of isolated 
galaxies, inflow which is often invoked to explain circumnuclear star formation 
and secular evolution (e.g.\ Kormendy \& Kennicutt 2004). 
Observations of many barred galaxies show evidence for dust lanes along the 
leading edges of the bar and these dust lanes likely trace the shocks and 
inflow driven by gravitational torques. 
The structure of circumnuclear dust within the semiminor axis of the bar 
can provide important information about the effectiveness of bar-driven 
mass transport. 

It is now possible to quantitatively study the connection between bars and 
their circumnuclear region due to a combination of three factors: near-infrared 
surface photometry of a large number of nearby galaxies, a relatively 
straightforward measure of bar strength $Q_b$ from near-infrared images 
(Buta \& Block 2001) through application of the gravitational torque method
of Combes \& Sanders (1981), and {\it HST}\ images of many of these 
galaxies.  
In this contribution I begin with a brief overview of the classification of 
circumnuclear dust structure. I then apply this system to a large sample of 
nearby galaxies and investigate correlations between bar strength and 
circumnuclear structure. 

\section{Circumnuclear Structure in Galaxies} 

Martini et al.\ (2003a) conducted an imaging survey of 123 nearby galaxies 
with the NICMOS and WFPC2 cameras on {\it HST}\ to study circumnuclear dust. 
These data were used to develop a purely empirical classification system 
based on common features in the dust distribution and without regard to 
either the larger scale or nuclear properties of the galaxy. This system is 
thus complementary to the subject of this conference as it is a dust 
classification scheme, rather than a {\it dust-penetrated} classification 
scheme. 
The classification system has six categories: 

\begin{itemize}
\item[]
Grand design (GD): Two dominant and coherent dust spirals 
\item[]
Tightly wound (TW): Coherent and large pitch angle dust spirals
\item[]
Loosely wound (LW): Coherent and small pitch angle dust spirals  
\item[]
Chaotic spiral (CS): Multiple, fragmented dust lanes implying the same 
sense of rotation. 
\item[]
Chaotic (C): Dust structure without a well-defined morphology 
\item[]
No Structure (N): No evidence for nuclear dust structure 
\end{itemize}
An example of each of these classes is shown in Figure~1. 

\begin{figure}[ht]
\centerline{\includegraphics[width=4.5in]{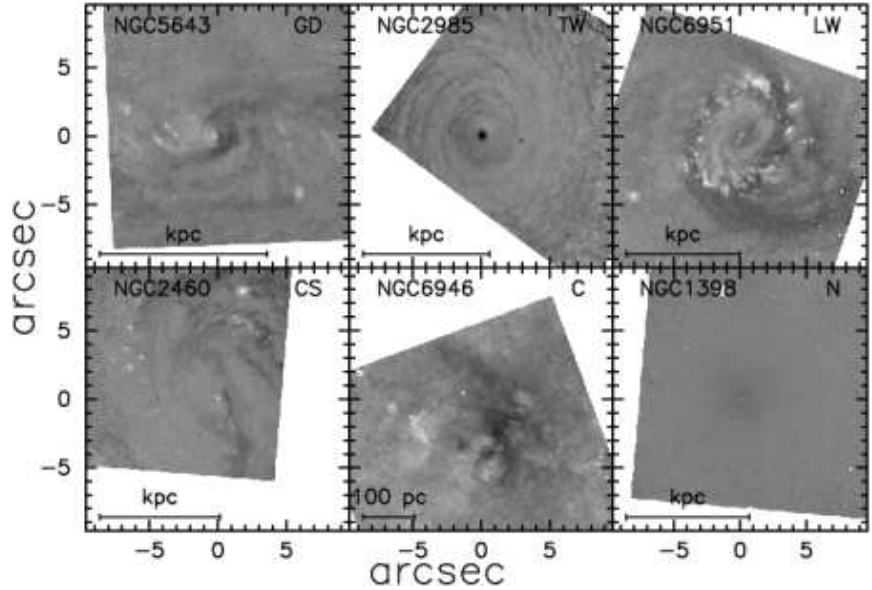}} \label{fig:class}
\caption{
$(V-H)$ color maps of prototypes for the six circumnuclear morphology classes 
proposed by Martini et al.\ (2003a) and reproduced from their Figure~3. 
}
\end{figure}

The initial sample was culled of all galaxies with 
$\upsilon > 5000$ km s$^{-1}$ and 
inclinations $R_{25} > 0.30$ and then each unbarred galaxy was matched with a 
barred galaxy of approximately the same morphological ($T$) type, blue 
luminosity, heliocentric velocity, inclination, and angular size. 
This resulted in an extremely well-matched set of 19 barred and 19 unbarred 
galaxies. 
The distribution of these 38 galaxies into the six circumnuclear dust 
classes is shown in Figure~2. This figure clearly demonstrates two connections 
between bars and their circumnuclear region: GD structure is only 
found in barred galaxies, while TW structure avoids barred galaxies 
(Martini et al.\ 2003b). 
In addition, GD structure often (but not always) connects to the dust 
lanes along the leading edges of the bar at larger scales. 

\begin{figure}[ht]
\centerline{\includegraphics[width=4.5in]{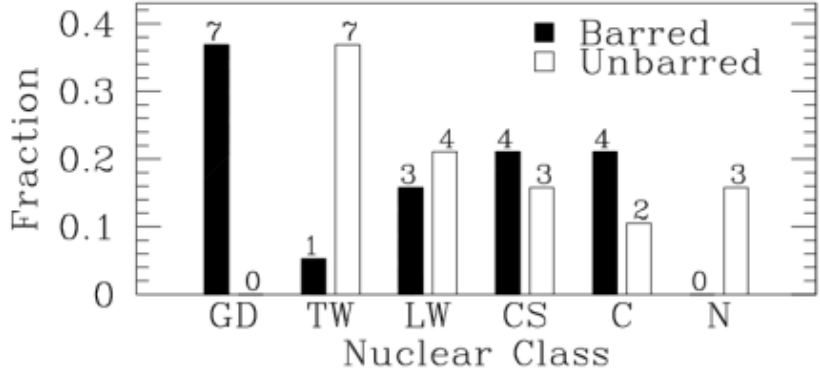}} 
\caption{
The frequency of the six circumnuclear classes in the sample of 19 barred 
and 19 unbarred galaxies studied by Martini et al.\ (2003b) and based on 
their Figure~2. GD structure is only found in barred galaxies, while 
TW structure appears to avoid barred galaxies. 
}
\end{figure}

These correlations also validate the classification system itself, as they 
indicate that the classification bins are connected to physically relevant 
quantities and are not simply lost in the dust. 
However, these results were limited by the absence of bar strength 
measurements for most of the sample. 
Although Martini et al.\ (2003a) collected data on whether or not a galaxy was 
barred from the literature, these data varied substantially in quality and 
were deemed too heterogeneous to investigate correlations between the 
circumnuclear morphology and bar strength. In the next section, I apply this 
classification system to {\it HST}\ observations of a large sample of galaxies 
with measured bar strengths. 

\section{Bar Strength and Circumnuclear Structure} 

The new sample described here was compiled from all galaxies with published bar 
strengths (Buta \& Block 2001; Laurikainen \& Salo 2002; Block et al.\ 2004), 
visible-wavelength images with {\it HST}, and inclinations $R_{25} < 0.30$. 
Galaxies with low signal-to-noise (S/N), unfavorable placement on the WFPC2 
detectors, or of type Scd ($T=6$) or later were barred from inclusion, 
although Scd galaxies are discussed separately below. 
The final sample contains 48 galaxies of type S0 to Sc. 

I used the structure map technique developed by Pogge \& Martini (2002) to 
identify circumnuclear morphology. For this application, 
structure maps are superior to color maps because they can be applied to the 
entire (larger) field of view of the WFPC2 camera and many galaxies 
only have WFPC2 images. Mathematically, structure maps are: 
\begin{equation}
S = \left[\frac{I}{I \otimes P}\right]\otimes P^{t}
\end{equation}
where $S$ is the structure map, $I$ is the original image, $P$ is the PSF, 
$P^{t}$ the transpose of the PSF, and $\otimes$ is the convolution operator. 
Structure maps effectively emphasize structures on the scale of the PSF and 
deemphasize larger-scale spatial variations. In all of the structure maps 
shown here, dusty regions are dark and emission regions, such as star formation 
knots, are bright. Figure~3 shows images, color maps, and structure maps 
of four representative galaxies. 

\begin{figure}[ht]
\centerline{\includegraphics[width=4.5in]{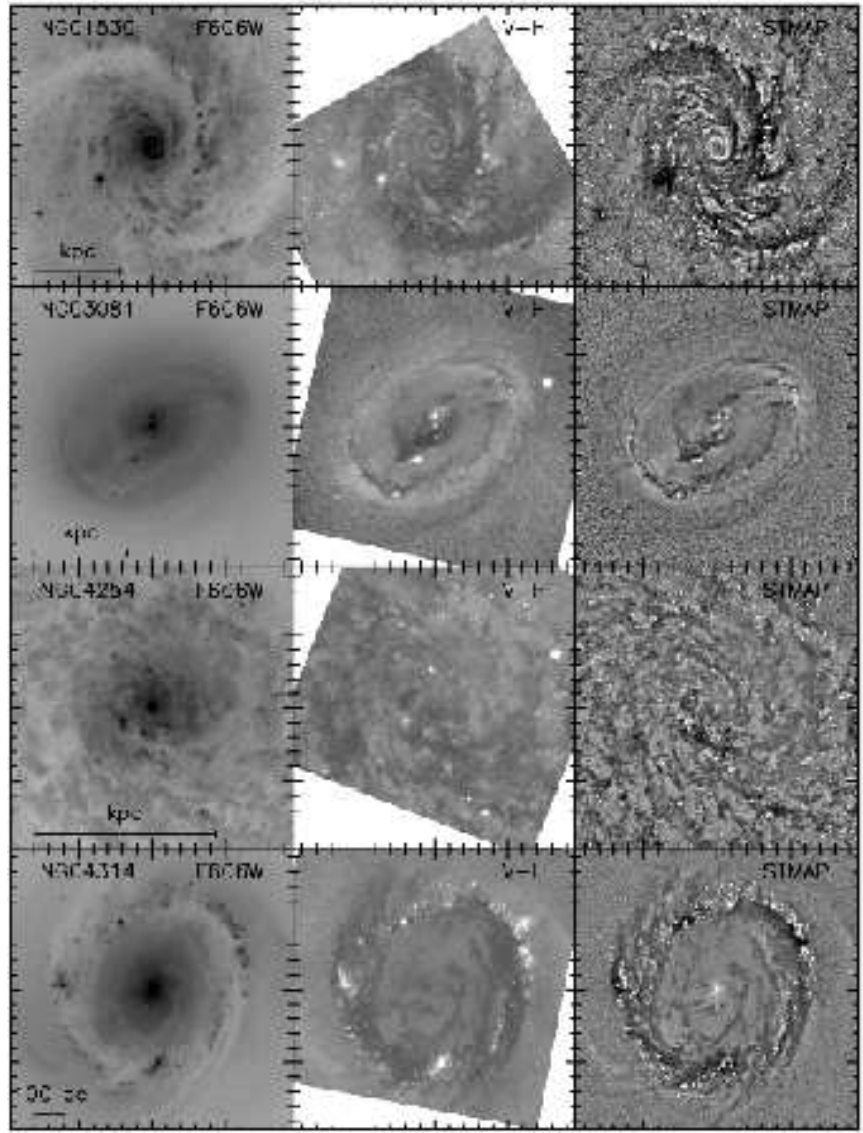}} 
\caption{
Comparison of $V$, $V-H$, and structure maps for NGC~1530, NGC~3081, NGC~4254, 
and NGC~4314. 
Both the color and structure maps effectively uncover dust features and 
star formation over a wider intensity range than the $V$ image, even with the 
log intensity scaling shown. Dusty regions are dark, while emission line 
regions are light. The structure and color maps are similar, although the 
structure maps place greater emphasis on features near the resolution limit. 
Each panel is $19''$ on a side and displays north up and east to the left. 
}
\end{figure}

\begin{figure}[ht] 
\centerline{\includegraphics[width=4.5in]{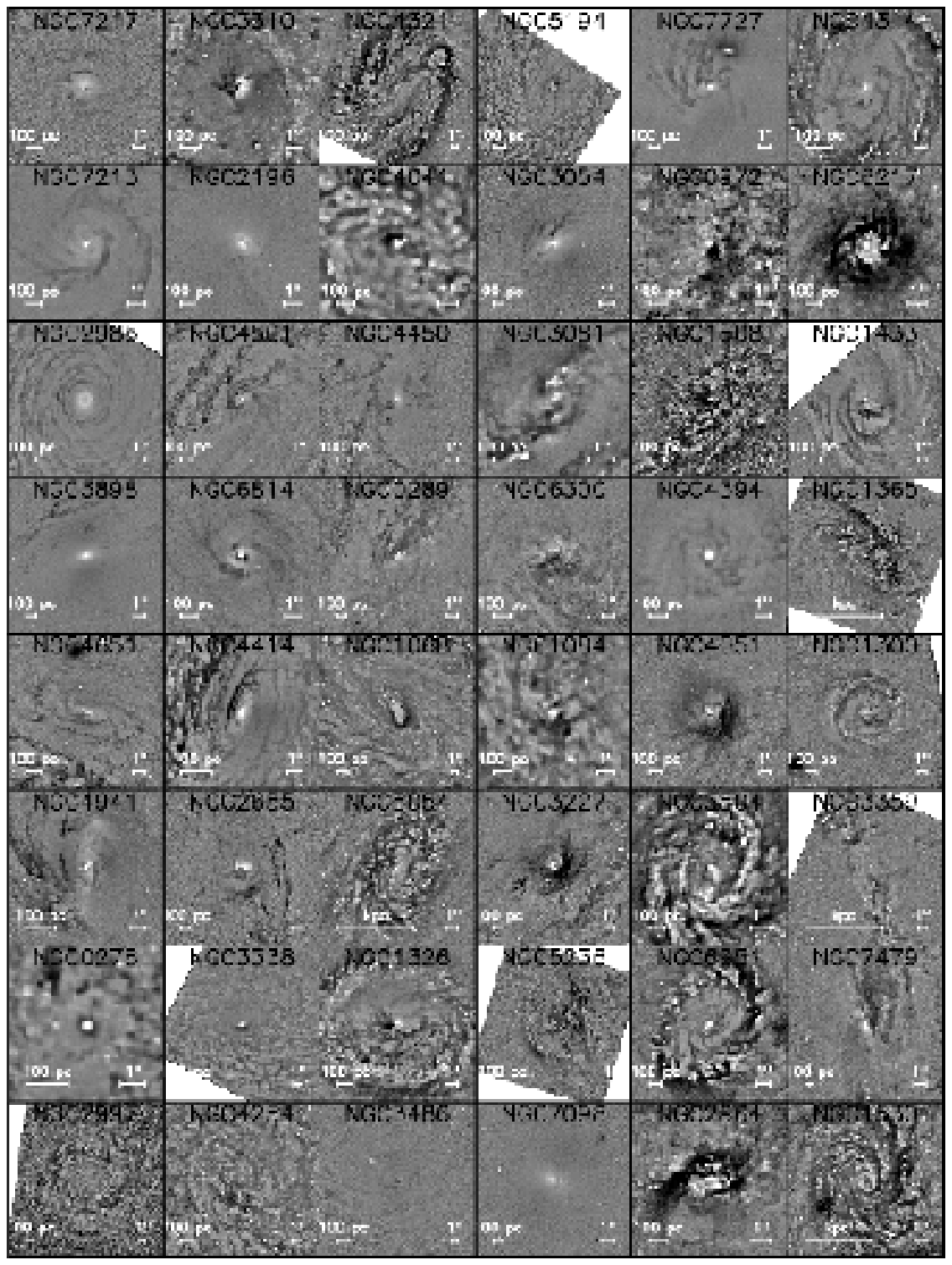}}
\caption{
Structure maps for 48 galaxies with measured bar strengths ordered such 
that bar strength increases downward and to the right. 
Each panel shows the inner 5\% of $D_{25}$ from the RC3 catalog. 
The projected size of a kpc or 100pc is shown to the lower left, while 
a $1''$ scale bar is shown at lower right. 
North is up and east is to the left. 
}
\end{figure}

The galaxies were classified with the same system described in the previous 
section. 
The only modification is that instead of the fixed angular size of $19''$ 
employed by Martini et al.\ (2003a), I have chosen to classify the sample 
within a fixed 5\% fraction of each galaxy's angular diameter $D_{25}$. 
This fractional size corresponds to a projected physical size range of 
$0.4\rightarrow2.4$kpc, while the projected physical size of the PSF is 
$2\rightarrow14$pc. 
Figure~4 presents structure maps of the central 5\% of the 48 galaxies. 
There are fourteen galaxies common to this sample and Martini et al.\ 
(2003a) and as a check they were reclassified without reference to the prior 
classification. Nine received the same classification, three switched
between the similar classes LW and CS, and only two (14\%) changed 
significantly: NGC~6300 (C$\rightarrow$GD) and NGC~4314 (LW$\rightarrow$GD). 

\begin{figure}[ht] 
\centerline{\includegraphics[width=4.5in]{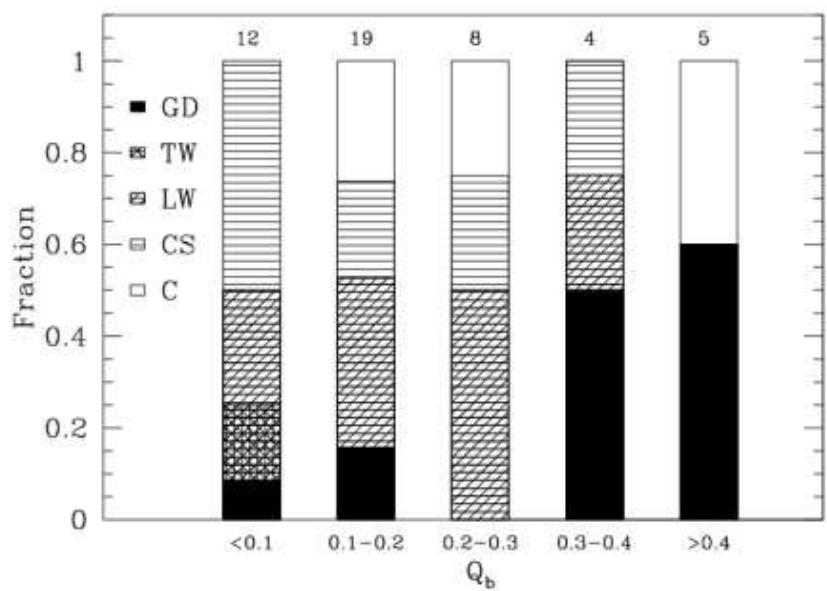}}
\caption{
Fraction of each NC class in bins of $Q_b$. Only a small fraction 
of weakly barred galaxies have GD structure, while it is present in 
5/9 galaxies with $Q_b > 0.3$. TW structure is only present in galaxies 
with $Q_b < 0.1$. The number of galaxies in each bin is shown above it. 
}
\end{figure}

Figure~5 shows the distribution of circumnuclear structure classes 
as a function of $Q_b$. The N class was not employed here because it was 
only populated by one galaxy (NGC~1398). 
Only four of 31 galaxies (13\%) with $Q_b < 0.2$ have GD structure, while 
it is present in five of nine (56\%) galaxies with $Q_b > 0.3$.  
More strongly barred galaxies are therefore more likely to have GD 
structure. The one GD galaxy with $Q_b < 0.1$ is NGC~6814, which was also 
classified GD in Martini et al.\ (2003a). This galaxy is listed as unbarred in 
the RC3, although it is classified as barred in the near-infrared. 

There are also no galaxies with TW structure and $Q_b>0.1$. Large pitch 
angle dust spirals are therefore not found in galaxies with a significant 
nonaxisymmetric component. 
While only two galaxies were classified as TW, the probability that both would 
have $Q_b < 0.1$ is 4\%. 
This result reinforces the suggestion of Martini et al.\ (2003b) that TW 
structure is only present in unbarred galaxies, and also supports recent 
simulation results (Maciejewski 2004). 

\subsection{Connection to larger-scale spirals: SB(s) and SB(r) galaxies} 

GD structure is preferentially found in galaxies with large $Q_b$ and 
in many cases appears to be the continuation of the dust lanes along 
the leading edges of large scale bars, dust lanes that models show are 
formed by strong bars (Athanassoula 1992). 
Another historical measure of bar strength is whether the large-scale spiral 
arms originate at the ends of a bar SB(s), from an inner ring at the radius 
of the bar SB(r), or are intermediate SB(rs). Observations show that dust 
lanes along the bar are a characteristic of SB(s) galaxies, rather than 
SB(r) galaxies (e.g.\ Sandage \& Bedke 1994). The presence of dust lanes 
suggests SB(s) bars should be strong, although hydrodynamic simulations 
find SB(s) spirals form with weak, fast bars and SB(r) spirals with strong, 
slow bars (Sanders \& Tubbs 1980). 

I have investigated the frequency of GD structure in all galaxies 
classified as SB in the RC3 (11 galaxies). 
This sample shows a correlation between GD structure and the connection between 
the bar and the large-scale spiral arms: Neither of the two SB(r) galaxies in 
this sample have GD structure, although it is present in $3/6$ SB(rs) galaxies 
and $2/3$ SB(s) galaxies. 
To test the connection between these classes and $Q_b$ for a larger sample, 
I have computed the mean $Q_b$ for all 45 galaxies classified as type SB(r): 
0.28 (15), SB(rs): 0.35 (12), and SB(s) 0.43 (18). 
On average SB(s) is thus the most strongly barred type and SB(r) the weakest. 
However, the SB(s) sample does include more galaxies with late $T$ type and 
large $Q_b$. 
If only galaxies with $T \leq 5$ are included (the range of the 
SB(r) sample), the mean value of $Q_b$ for the SB(s) class decreases to 0.35 
(11 galaxies). 
It would be valuable to revisit the SB(r)/SB(s) classification with 
near-infrared images.

\subsection{Connection to global properties} 

I have also used this sample to investigate if circumnuclear structure 
depends on $T$ type, luminosity, or distance. The sample was divided 
into early ($T\leq1$; 10 galaxies), intermediate ($T=2,3$; 16), and late 
($T=4,5$; 22) type bins. There is a gradual increase in the fraction of 
LW structure ($1/10\rightarrow4/16\rightarrow10/22$) and an approximately 
corresponding 
decline in the fraction of C structure ($4/10\rightarrow3/16\rightarrow3/22$). 
No change is observed in the fractional distribution of the remaining classes. 
The increase in the LW fraction at the expense of the C fraction 
suggests an increase in the fraction of galaxies that have circumnuclear 
gaseous disks, as a circumnuclear disk is required for spiral dust lanes to 
form (types GD, TW, LW, CS). 
There are no obvious trends with distance or luminosity, although this is not 
surprising because these galaxies span a relatively narrow range in luminosity 
and distance. This does indicate that the approximately factor of five 
range in the projected physical size of the kernel does not affect these 
results. 

\subsection{Very late-type galaxies} 

Figure~6 displays the nine Scd galaxies with measured $Q_b$ in the {\it HST}\ 
archive, although only two meet the standards of the main sample. 
These two galaxies are both of type C, as are all but two of the other
(lower S/N) galaxies. 
This may indicate that there is a dramatic increase in the fraction of 
galaxies with chaotic structure at very late $T$ type, or it may only reflect 
the importance of high S/N for accurate classification of circumnuclear dust 
structure. 

\begin{figure}[ht] 
\vskip.2in
\centerline{\includegraphics[width=4.5in]{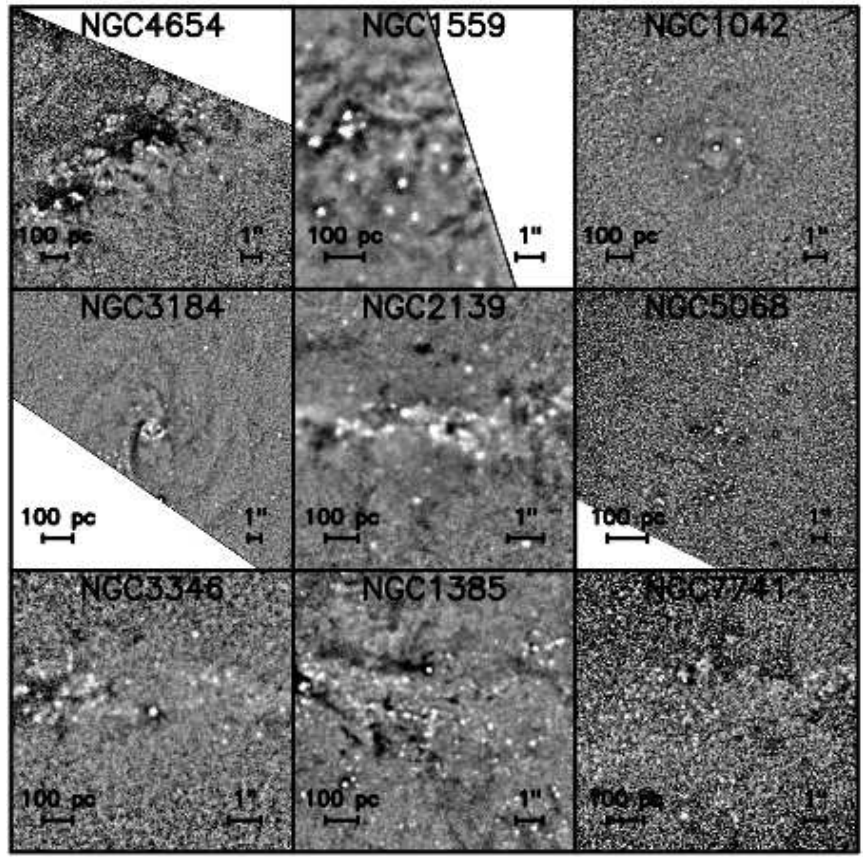}}
\caption{
Same as Figure~4 for late-type galaxies ($T=6$). Only NGC~2139 and NGC~3184 
meet the same S/N and position requirements as the main sample. The 
structures in most panels are dominated by star forming regions, rather 
than dust. 
}
\end{figure}

\section{Discussion and Summary}

I have computed structure maps from {\it HST}\ data for 48 galaxies with 
measured bars strengths $Q_b$. 
These data clearly shown that the fraction of galaxies with GD structure 
increases sharply for stronger bars, while TW structure is only found in the 
most axisymmetric galaxies. 
GD structure is more common in galaxies classified as type SB(s), which are 
commonly observed to have dust lanes along the leading edges of their bars. 
Measurement of the mean $Q_b$ for SB(s) and SB(r) galaxies shows that SB(s) 
galaxies generally have stronger bars. 
GD structure, SB(s) structure, and dust lanes along bars are therefore all 
correlated with bar strength. 
The fraction of galaxies with nuclear dust spirals increases at later $T$ type, 
perhaps indicating an increase in the fraction of galaxies with circumnuclear 
gaseous disks. 
There is some evidence that this trend reverses at type Scd with an increase 
in the fraction of C structure, although the available data for these 
galaxies are significantly poorer quality. 

\begin{acknowledgments}
I was supported during the course of this work by a Clay Fellowship at the 
Harvard-Smithsonian Center for Astrophysics. 
I acknowledge the support of the American Astronomical Society and the 
National Science Foundation in the form of an International Travel Grant, 
which enabled me to attend this conference.
Support for this work was also provided by NASA through grant number AR-9547 
from the Space Telescope Science Institute, which is operated by the 
Association of Universities for Research in Astronomy, Inc., under NASA 
contract NAS5-26555. 
\end{acknowledgments}


\begin{chapthebibliography}{1}

\bibitem{athanassoula92}
Athanassoula, E. 1992, MNRAS, 259, 345

\bibitem{block04}
Block, D.L. et al.\ 2004, AJ, {\it press}, (astro-ph/0405227) 

\bibitem{buta01}
Buta, R. \& Block, D.L. 2001, ApJ, 550, 243

\bibitem{buta04}
Buta, R., Laurikainen, E., \& Salo, H. 2004, AJ, 127, 279

\bibitem{combes81}
Combes, F. \& Sanders, R.H. 1981, A\&A, 96, 164 

\bibitem{kormendy04}
Kormendy, J. \& Kennicutt, R.C. 2004, ARA\&A, {\it in press} 

\bibitem{laurikainen02}
Laurikainen, E., \& Salo, H. 2002, MNRAS, 337, 1118 

\bibitem{maciejewski04}
Maciejewski, W. 2004, in Carnegie Observatories Astrophysics Series, Vol. 1: 
Coevolution of Black Holes and Galaxies, ed. L. C. Ho (Pasadena: Carnegie 
Observatories, \\
http://www.ociw.edu/ociw/symposia/series/symposium1/proceedings.html)

\bibitem{martini03a}
Martini, P., Regan, M.W., Mulchaey, J.S., \& Pogge, R.W. 2003, ApJS, 146, 353

\bibitem{martini03b} 
Martini, P., Regan, M.W., Mulchaey, J.S., \& Pogge, R.W. 2003, ApJ, 589, 774

\bibitem{pogge02}
Pogge, R.W. \& Martini, P. 2002, ApJ, 569, 624 

\bibitem[Sandage \& Bedke (1994)]{sandage94}
Sandage, A., \& Bedke, J. 1994, The Carnegie Atlas of Galaxies
(Publ. 638; Washington, DC: Carnegie Inst. Washington) 

\bibitem{sanders80}
Sanders, R.H. \& Tubbs, A.D. 1980, ApJ, 235, 803

\end{chapthebibliography}
\end{document}